\documentclass[tighten,preprint2,flushrt]{aastex}
\usepackage{float}
\usepackage{multirow}
\usepackage{subfigure}
\usepackage{color}
\usepackage{color,soul}
\usepackage{rotating}
\newcommand{\myemail}{zouhu@nao.cas.cn}

\begin{document}
\title{Project Overview of the Beijing-Arizona Sky Survey}
\author{Hu Zou\altaffilmark{1}, Xu Zhou\altaffilmark{1}, Xiaohui Fan\altaffilmark{2}, Tianmeng Zhang\altaffilmark{1}, Zhimin Zhou\altaffilmark{1}, Jundan Nie\altaffilmark{1}, Xiyan Peng\altaffilmark{1}, Ian McGreer\altaffilmark{2}, Linhua Jiang\altaffilmark{3}, Arjun Dey\altaffilmark{4}, Dongwei Fan\altaffilmark{1}, Boliang He\altaffilmark{1}, Zhaoji Jiang\altaffilmark{1},  Dustin Lang\altaffilmark{5}, Michael Lesser\altaffilmark{2}, Jun Ma\altaffilmark{1}, Shude Mao\altaffilmark{6,1,7},  David Schlegel\altaffilmark{8}, Jiali Wang\altaffilmark{1}} 
\altaffiltext{1}{Key Laboratory of Optical Astronomy, National Astronomical Observatories, Chinese Academy of Sciences, Beijing 100012, China; \myemail}
\altaffiltext{2}{Steward Observatory, University of Arizona, Tucson, AZ 85721}
\altaffiltext{3}{Kavli Institute for Astronomy and Astrophysics, Peking University, Beijing 100871, China}
\altaffiltext{4}{National Optical Astronomy Observatory, Tucson, AZ 85719}
\altaffiltext{5}{David Dunlap Institute, University of Toronto, Toronto, Canada}
\altaffiltext{6}{Department of Physics and Tsinghua Center for Astrophysics, Tsinghua University, Beijing 100086, China}
\altaffiltext{7}{Jodrell Bank Centre for Astrophysics, University of Manchester, Manchester M13 9PL, UK}
\altaffiltext{8}{Lawrence Berkeley National Labortatory, Berkeley, CA 94720}

\begin{abstract} 
The Beijing-Arizona Sky Survey (BASS) is a wide-field two-band photometric survey of the Northern Galactic Cap using the 90Prime imager on the 2.3 m Bok telescope at Kitt Peak. It is a four-year collaboration between the National Astronomical Observatory of China and Steward Observatory, the University of Arizona, serving as one of the three imaging surveys to provide photometric input catalogs for target selection of the Dark Energy Spectroscopic Instrument (DESI) project. BASS will take up to 240 dark/grey nights to cover an area of about 5400 deg$^2$ in the $g$ and $r$ bands. The 5$\sigma$ limiting AB magnitudes for point sources in the two bands,  corrected for the Galactic extinction, are 24.0 and 23.4 mag, respectively. BASS, together with other DESI imaging surveys, will provide unique science opportunities that cover a wide range of topics in both Galactic and extragalactic astronomy.
\end{abstract}


\section{Introduction}

The Sloan Digital Sky Survey \citep[SDSS;][]{yor00}, with its unique combination of wide-field spectroscopic and imaging surveys, has made revolutionary discoveries in astrophysics and cosmology, from the studies of dark energy, the most distant quasars, to near-field cosmology, Galactic structure and the cool brown dwarfs. The SDSS imaging survey was carried out more than a decade ago and only reached r$\sim$22 mag, which is sensitive to low-redshift galaxies, luminous quasars, and the tip of iceberg of the Galactic halo substructure. There is a gap between the imaging data currently available and deep imaging data that future surveys will provide (e.g. the Large Synoptic Survey Telescope, or LSST). This is more obvious in the Northern hemisphere, because most new imaging surveys that are being carried out or planned are in the south, such as SkyMapper, Dark Energy Survey (DES) , and LSST \citep{des05,kel07,lss09}. 

Deeper images are also needed to fully explore some existing spectroscopic data. SDSS is an example. The SDSS-III/BOSS survey has obtained a large number of spectra over 9400 deg$^2$, including about 0.25 million stars, 1.4 million galaxies, and 0.29 million quasars \citep{ala15}. SDSS-IV/eBOSS (2014-2020) will add another 0.6 million galaxies at $0.6 < z < 1$ and 0.75 million quasars at $z > 0.9$ \citep{daw16}. Although SDSS-I spectroscopy is relatively shallow, BOSS and eBOSS have been pushing to the limit of the SDSS imaging data \citep{daw13,daw16}. The SDSS images are no longer sufficient for future spectroscopic target selection. Without deeper images, we are not able to measure accurate photometry, morphology, and detailed information of the galactic environment that are needed to study galaxy properties at lower luminosity and higher redshift. 

The Dark Energy Spectroscopic Instrument (DESI) project will conduct a next generation cosmological spectroscopic survey, covering about 14,000 deg$^2$ in 2018--2022, using the 4m Mayall telescope on Kitt Peak with a 5000 fiber spectroscopic system \citep{fla14}. It will accurately measure the expansion rate and structure growth of the universe by measuring spectroscopic redshift of about 34 million objects, including five populations: Bright Galaxies (BGs), Luminous Red Galaxies (LRGs), Emission Line Galaxies (ELGs), low-redshift quasars, and high redshift Lyman-$\alpha$ forest quasars. These targets will be selected using three optical bands ($g$, $r$, and $z$) and the first two infrared bands W1 and W2 of the Wide-field Infrared Survey Explorer \citep[WISE;][]{wri10}. 

The Beijing-Arizona Sky Survey (BASS\footnote{\url{http://batc.bao.ac.cn/BASS/}}) is a collaborative project between National Astronomical Observatory of China (NAOC) and the Steward Observatory of the University of Arizona (UA). A team of Chinese astronomers,  has joined DESI through the BASS project as a Regional Participation Group.  BASS performs the $g$ and $r$-band imaging in the Northern Galactic cap using Steward Observatory's 2.3m Bok telescope on Kitt Peak, Arizona. The survey depth will be about 1--2 mag deeper than the depth of the SDSS imaging survey. BASS, Dark Energy Camera Legacy Survey \citep[DECaLS;][]{blu16}, and MOSAIC z-band Legacy Survey \citep[MzLS;][]{sil16} constitute the DESI imaging surveys.  In addition to providing spectroscopic targets for DESI, these optical imaging surveys together with the WISE infrared survey will allow us to study Galactic structures, galaxy clusters, AGN evolution, high redshift quasars, large scale structures, and so on. 

This paper provides an overview of the BASS project, which is organized as follows. Section \ref{sec-survey} briefly describes the survey basics, including the facilities and survey footprint. Section \ref{sec-science} shows some scientific capabilities with BASS and other DESI imaging data. The observing strategy and progress are presented in Section \ref{sec-obs}. The data release plan is briefly mentioned in Section \ref{sec-release}. Section \ref{sec-summary} gives a summary.

\section{The Survey} \label{sec-survey}
BASS was initiated by NAOC and UA, with participants from Chinese universities and other members of  the DESI collaboration. The project was proposed in the middle of 2014 and approved as a key project of the Chinese Telescope Access Program (TAP\footnote{\url{http://info.bao.ac.cn/tap/?q=keyprojects}}) in September of that year. BASS uses the 2.3 m Bok telescope to obtain deep imaging data in the $g$ and $r$ bands with 5$\sigma$ depths of 24.0 and 23.4 mag, respectively. The depths are corrected for Galactic extinction for targeting extragalactic sources. The observations started in 2015 and will last four years from 2015 through 2018. Table \ref{tab-summary} gives a summary of the survey. 

\begin{table}[!h]
\centering
\caption{BASS Survey Summary} \label{tab-summary}
\begin{tabular}{ll}
\hline
\hline
Telescope  & 2.3 m Bok Telescope \\
Site & Kitt Peak (altitude 2071m) \\
Corrected focal ratio & f/2.98 \\
Corrected focal length & 6.83 m \\
Camera & 90Prime \\
Detector & four 4k$\times$4k CCDs \\
Pixel scale & 0\arcsec.454 (15 $\mu$m) \\
FoV & 1\arcdeg.08$\times$1\arcdeg.03 \\
Filter & SDSS $g$, DES $r$ \\
Survey area & 5400 deg$^2$ \\
Number of exposures & 3 \\
Full depth (mag; 5$\sigma$) & $g = 24.0$, $r = 23.4$ \\
Period & 2015--2018 \\
\hline
\end{tabular}
\end{table}

\subsection{Telescope and Instrument}
The Steward Observatory Bok telescope is an equatorial mounting telescope with a diameter of 2.3 meters. The corrected primary focal ratio is f/2.66 and the telescope pointing error is about 3\arcsec. It is located on Kitt Peak and has a typical seeing of 1\arcsec.5 \citep{zou15a}. 

BASS uses the prime-focus imager, 90Prime, with a corrected focal ratio of f/2.98. The optical design includes a four-element corrector and six position filter wheel. The focal plane array is a mosaic of four 4k by 4k CCDs that have been processed for back illumination by the UA Imaging Technology Laboratory. These CCDs were optimized for the ultraviolet response and built for the South Galactic Cap $u$-band Sky Survey \citep[SCUSS;][]{zho16} \footnote{\url{http://batc.bao.ac.cn/Uband/}}. The quantum efficiencies in the $g$ and $r$ bands are higher than 80\% (see Figure \ref{fig-filter}).  The edge-to-edge field of view (FoV) is about 1\arcdeg.08$\times$1\arcdeg.03, and the pixel scale is 0\arcsec.454. The gaps between CCDs along right ascension and declination are 2\arcmin.8 and 0\arcmin.9, respectively. The average gain is about 1.5 e/ADU and the average readout noise is about 8 e. 

\subsection{Filters}
\begin{figure}[!ht]
	\includegraphics[width=\columnwidth]{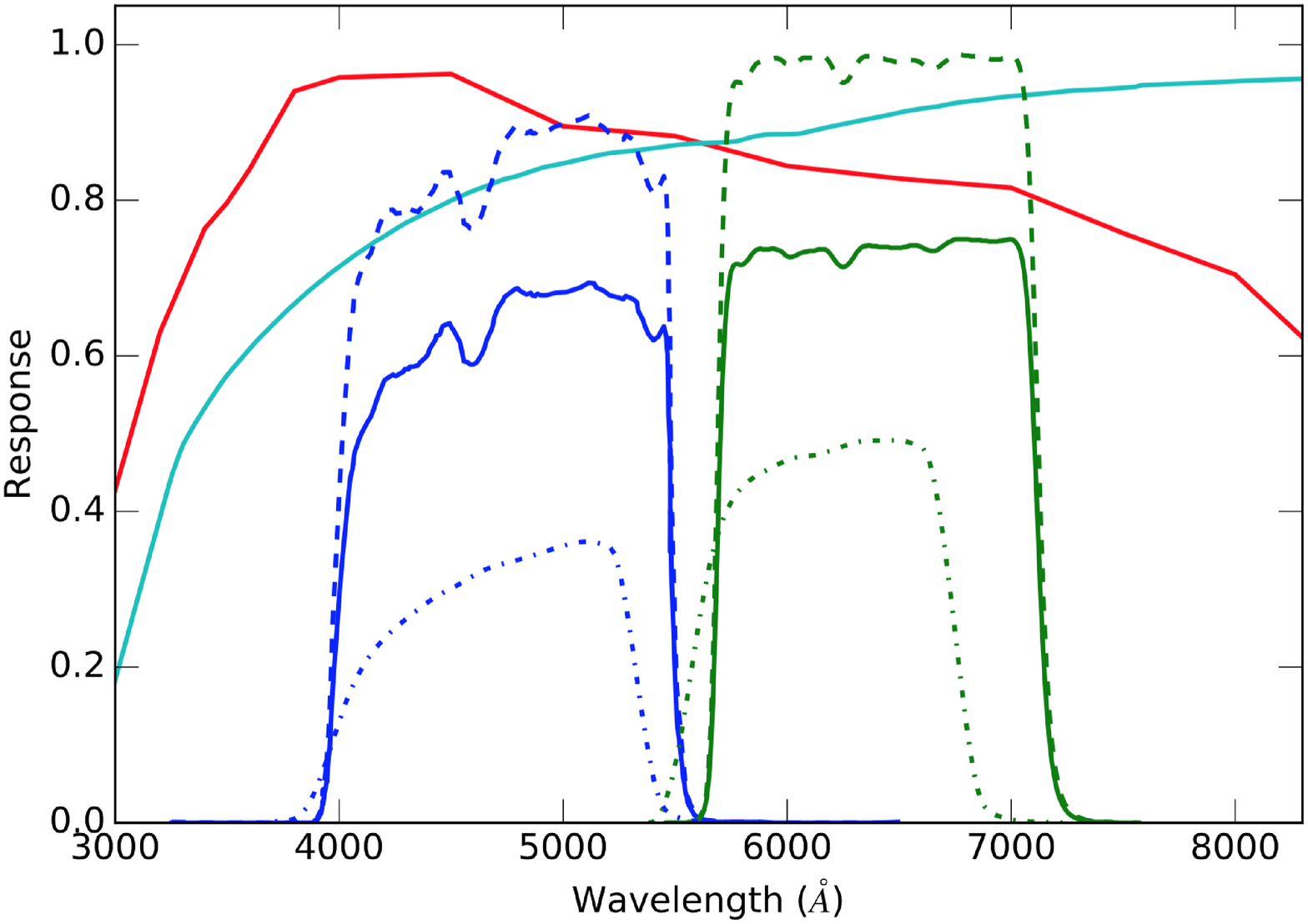} 
	\caption{Filter responses for the BASS $g$ (blue) and $r$ (red) bands. The red solid curve is the CCD QE and the cyan curve presents the atmospheric transmission on Kitt Peak. The dashed profiles are the transmission curves of these two filters. The solid profiles are the total responses, including the CCD QE and atmospheric extinction. The dash-dotted curves show the $g$ and $r$-band responses used by the SDSS.}  \label{fig-filter}
\end{figure}

The filters are required to be similar to those used by the DES and DECaLS to ensure a uniform photometric system  for DESI target selection. For BASS $g$ band, we are using an existing SDSS $g$ filter deployed on the Bok telescope. This filter is similar to the DES $g$ band. The BASS $r$ filter was newly purchased for this project. It is almost identical to the $r$ filter used by the DES and DECaLS, and is significantly different from the SDSS $r$ filter. The cut-on and cut-off wavelengths are 5680 and 7110 \AA, respectively. The absolute transmission is larger than 95\%. Figure \ref{fig-filter} shows the response curves of the BASS $g$ and $r$ filters. The response curves shown include the filter transmission, CCD quantum efficiency (QE), and the atmospheric extinction at airmass of 1.0. The atmospheric extinction coefficients come from the Kitt Peak National Observatory and molecular absorptions are not considered. The effective wavelength and bandwidth for the $g$ band are about 4776 and 848 {\AA}, and for the $r$ band are about 6412 and 833 {\AA}. The BASS $r$ filter is more than 200 {\AA} redder and has a broader passband than the SDSS $r$ filter. 

\subsection{Footprint and Tiling Strategy}
\begin{figure}[!ht]
	\includegraphics[width=\columnwidth]{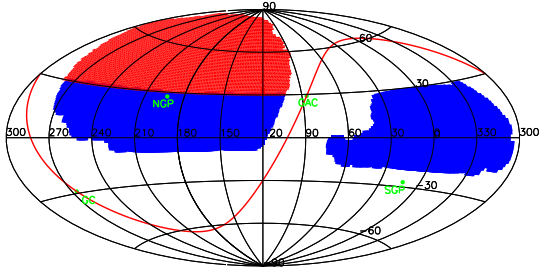} 
	\caption{The footprint of the DESI imaging surveys in the Aitoff projection centered at ($\alpha=120\arcdeg$, $\delta = 0\arcdeg$). The blue region shows the DECaLS coverage, and the red one shows the coverage of BASS and MzLS. The red curve represents the Galactic plane. The green points with labels show the positions of the Galactic center (GC), Galactic anticenter (GAC), north Galactic pole (NGP) and south Galactic pole (SGP). \label{fig-footprint}} 
\end{figure}

The DESI imaging Survey will cover a total area of about 14,000 deg$^2$ visible from the Northern hemisphere. The footprint is approximately enclosed  by $-18\arcdeg < \delta < +84\arcdeg$ (declination) and $|b| > 18\arcdeg$ (Galactic latitude), where regions with larger Galactic extinctions are excluded. The general DESI footprint is shown in Figure \ref{fig-footprint}. The blue region along the equator is the DECaLS footprint of about 9000 deg$^2$. The BASS coverage shown in red is located in the north of the northern Galactic cap with $\delta > 30\arcdeg$. The area is about 5400 deg$^2$. The MzLS will cover the same area but in the $z$ band.  

The BASS footprint will be scanned in three passes (or exposures). Each exposure is required to have similar depths.  The tiling strategy is designed in consideration of: (1) filling factor of the footprint should be larger than 90\%; (2) common area covered by all three exposures should be as large as possible.  The tiles are evenly-spaced over the footprint. Each tile has the size of the camera FoV. There are 5494 tiles for each pass. We make an offset of about $\pm\frac{1}{4}$ FoV to make sure that most of the CCD gaps are covered by three exposures. The offset is partly determined by assuming that the telescope pointing error is less than 25\arcsec. As a result, about 90.5\% of the footprint is covered by three exposures. The percentages of the area covered by two and one exposures are 9.3\% and 0.2 \%, respectively. 

\section{Observation} \label{sec-obs}
\subsection{Observing Strategy}
The BASS observing strategy is quite similar to those used for other DESI imaging surveys (K. Burleigh et al. 2017 in preparation). The three passes are observed under different weather conditions:
\begin{itemize}
\item Pass 1: photometric condition with seeing $<$ 1\arcsec.7;
\item Pass 2: either photometric condition or seeing $<$ 1\arcsec.7, or Pass 1 condition if  Pass 1 is completed;
\item Pass 3: non-photometric condition and seeing $>$ 1\arcsec.7, or any other condition if both Pass 1 and Pass 2 are completed. 
\end{itemize}
Pass 1 will be used for accurate internal photometric calibration.  The above observing strategy is also beneficial for the selection of variable objects such as variable stars and quasars \citep{zou15b}. 

We choose tiles to observe nightly according to the following criteria: 
\begin{itemize}
\item far away from the Moon ($> 60\arcdeg$);
\item close to its best airmass ($\Delta$AM $<0.1$);
\item priority given to declination below 60\arcdeg ($\delta < 60$\arcdeg);
\item priority given to Galactic extinction less than 0.1 (E$(B-V) < 0.1$).
\end{itemize}
BASS tiles with $\delta <60\arcdeg$ have high priorities, selected for synergy with MzLS.  Tiles with high Galactic extinction have lower priorities, because much longer exposure time is required to reach the same imaging depth. The $g$-band observations are required to be performed in dark time, and the $r$-band imaging is performed in dark or grey time. During our observations, we generate observing scripts in real time, based on weather conditions and sky brightness. Observing overhead comes from the CCD read-out, file transferring, telescope moving, and other time consumption. 

\subsection{Diagnostic plots}
\begin{figure}[!ht]
	\includegraphics[width=\columnwidth]{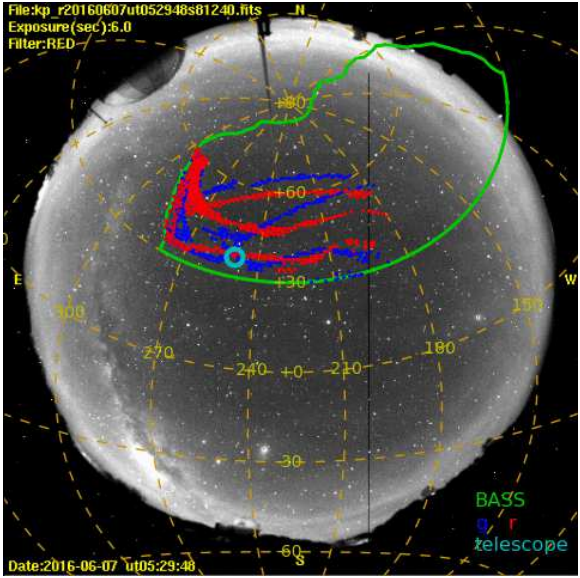}
	\caption{An updated image taken by the all-sky camera installed on Kitt Peak. The green envelope shows the BASS footprint. The points (blue for $g$ band and red for $r$ band) are the tiles to be observed during the night. The telescope pointing is over-plotted in cyan. The yellow dashed lines display the Equatorial grid. \label{fig-allskycam}}
\end{figure}

A number of diagnostic plots are presented on our wiki site for purpose of assessing image quality and remote monitoring. We have updated  all-sky camera image on Kitt Peak, which is overlaid with the planned tiles to be observed during the night (see Figure \ref{fig-allskycam}). The all-sky camera is helpful for observers to determine when to change to different passes according to the weather conditions. 

\begin{figure}[!ht]
	\includegraphics[width=\columnwidth]{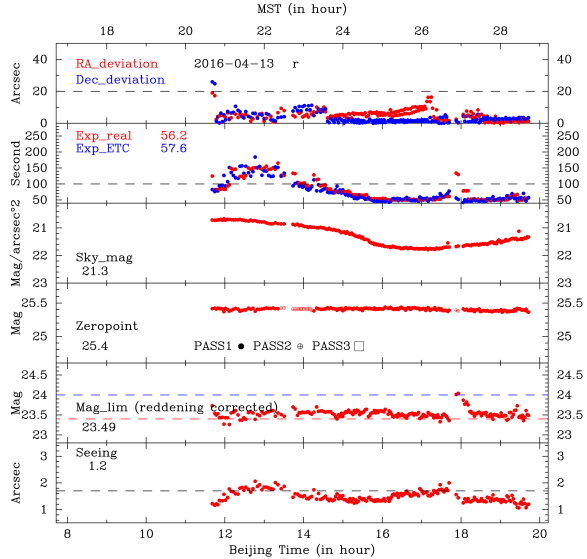}
	\caption{The telescope pointing error and image quality varying with time during a whole night.  From top to bottom, the panels show the parameter variations for R.A. (red) and decl. (blue) errors in arcsec of the telescope pointing, real-time exposure time (red) and exposure time in second given by the ETC (blue), sky brightness in mag/arcsec$^2$, zero point in mag for 1e/s, full depth in mag, and seeing in arcsec. The number shown in each panel gives the parameter value of the last point. The top x axis is the Mountain Standard Time in hour, and the bottom one gives the Beijing time. Different passes are shown in the third panel with different symbols. The dashed line in the first panel gives the upper limit of 20\arcsec. The one in the second panel indicates the typical single exposure time of 100 s. The blue and red dashed lines in the fifth panel show the defined depths of $g$ and r bands, respectively. The dashed line in the last panel gives the seeing cut of 1\arcsec.7.  \label{fig-quality}}
\end{figure}

During our observations, we carry out a quick data reduction in real time, including image detrending, astrometric and photometric calibration. We then estimate the telescope pointing errors in R.A. and decl., seeing, sky brightness, photometric zero points, and magnitude limits. An example is shown in Figure \ref{fig-quality}, including the variations of these parameters during the whole night. During our observations, these plots will alert us to change the scripts according to our observing strategy. We require that the telescope pointing error is less than 20{\arcsec} and the estimated magnitude limit is 0.1 mag deeper than the expected depths. The depths are also estimated by an exposure time calculator (ETC), which is same as used in all DESI imaging surveys (K. Burleigh et al. 2017 in preparation). The ETC dynamically changes the exposure time according to the current seeing, sky brightness, zero point, and Galactic extinction. We adopt the reddening map from \citet{sch98} and the extinction coefficients of $A_g = 3.303$ and $A_r = 2.285$ in the two bands. The typical exposure time for $g$ and $r$ band is about 300s, i.e. 100s for each pass. The minimum and maximum single exposures are set to be 50s and 250s, respectively. The magnitude limit shows in the fourth panel of Figure \ref{fig-quality} is the full depth assuming that three exposures are the same with a correction for the Galactic extinction. The full depth is approximately 0.6 mag deeper than single exposures. 

\subsection{Observation Progress}
Based on the average night duration of $\sim$9.6 hours, observing overhead, tiling strategy, and a fraction of good weather of 65\%, we estimate that a total of 240 nights would be needed to complete the BASS survey. This also include a 10\% loss for instrument failure and other contingency.  The BASS observations are carried out in the first semester of each year, from January 2015 through June 2018. 

\begin{figure*}[!ht]
\centering
	\includegraphics[width=1.0\textwidth]{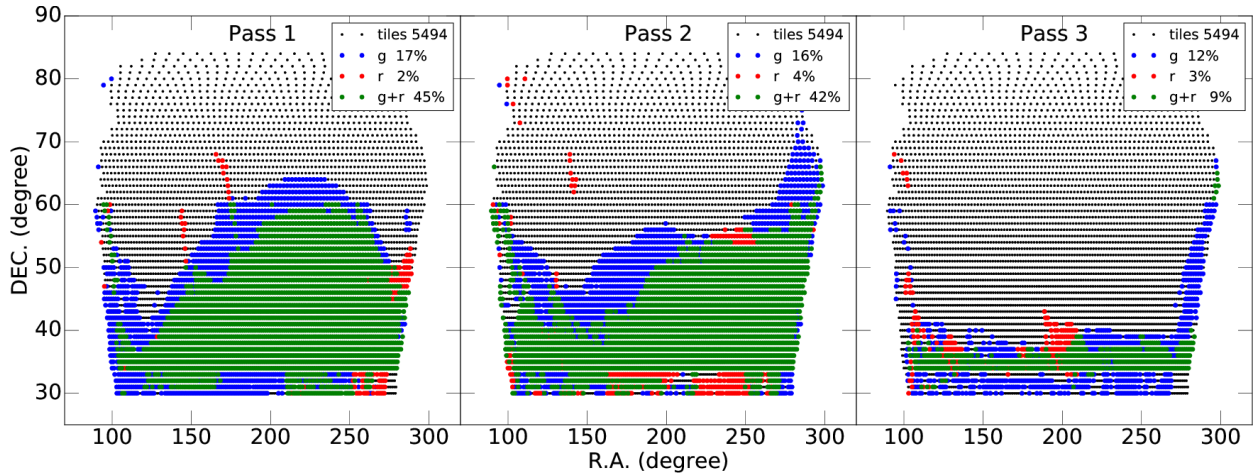} 
	\caption{The observation progress in the past two years of 2015 and 2016. From left to right, the panels are progresses for Pass 1, Pass 2, and Pass 3. The bad-quality data taken in 2015 are removed. The black point shows the BASS tiles of one pass. The blue and red points present the tiles that have been observed for only $g$ band and only $r$ band, while the green points show the tiles observed with both filters. The percentages in the legends show the fractions of tiles that have been observed. \label{fig-progress}} 
\end{figure*}

We observed 55 and 89 nights in 2015A and 2016A, respectively. The average fraction of good weather that is suitable for BASS is about 68\%. The tiles that do not meet our depth requirement will be observed again in following observing runs. A total of about 38\% of $g$-band tiles were completed in 2015, while only 6\% of $r$-band imaging was finished due to the late arrival of the new filter. The ETC in 2015 overestimated the imaging depths and thus underestimated the exposure time. We have made a number of updates in 2016 on telescope control softwares and observing tools, which improved the telescope performance and observing efficiency. Observations in 2016 ran smoothly. Figure \ref{fig-progress} presents the survey progress in the past two years, where low-quality data are discarded.  In total, we have completed about 41\% of all BASS observations in 2015 and 2016.  

\section{Data Release Plan} \label{sec-release}
The early data were mainly reduced by the SCUSS pipeline, and was released publicly in December 2015 \footnote{\url{http://batc.bao.ac.cn/BASS/doku.php?id=datarelease:edr:home}}. SCUSS used the same telescope and camera and imaged an area of 5000 deg$^2$ of the south Galactic cap in the $u$ band. The astrometric and photometric solutions are tied to the Forth US Naval Observatory CCD Astrograph Catalog \citep[UCAC4;][]{zac13} and SDSS, respectively. The photometry is performed on single-epoch images with prior information of source positions and shapes from the SDSS catalogs. For more details about the project, data reduction, and data release, we can refer to all related papers \citep{zho16,zou15a,zou16}. Note that the 2015 data suffered from problematic A/D converters that have introduced additional errors in the raw images. These electronic elements were replaced and the new ones performed well in 2016. 

New data reduction pipelines have been developed for the First Data Release (DR1). For flat-fielding in DR1, we combine both the dome flats and super sky flats. This is different from the early data reduction that only used the super sky flats. New intra-CCD and inter-CCD crosstalk coefficients are derived. In addition, CCD artefacts, such as black cores of saturated stars due to integer overflow, satellite tracks, cosmic rays, and the bleeding trails of saturate stars, are masked. The astrometry is tied to SDSS/2MASS. The photometric solutions are derived based on the point source catalogs from Pan-STARRS1 \citep{cha16}. We perform photometry on single-epoch images and generate co-added catalogs by cross-matching the magnitude measurements from corresponding single-epoch catalogs. More details about DR1 can be referred to the data release paper \citep{zou17}. 

We expect to release future new data once a year. New BASS data and MzLS z-band data will be integrated and the data reduction pipelines will be also continuously updated. The following list includes some possible updates in future data releases:
\begin{itemize}
\item reprocessing the MzLS data with the BASS pipelines;
\item astrometry tied to Gaia;
\item refining the astrometry at the catalog level;
\item internal photometric calibration;
\item source detections on stacked images;
\item model photometry with exponential, deVauculeurs, and composite profiles;
\item forced model photometry with same shape parameters;
\item merging WISE data.
\end{itemize}

\section{Scientific Capabilities} \label{sec-science}
\subsection{The Milky Way Halo}
The SDSS revolutionized our view of the Milky Way (MW), especially its halo and environment. From the SDSS imaging data, people have found numerous stellar halo streams \citep{new03,yan03,gri09} and have doubled the number of dwarf galaxies \citep{wil10}. BASS, MzLS, and DECaLS will reach more than a factor of two further into the halo and increase the volume by nearly an order of magnitude. 

\subsubsection{Galactic structure through proper motion}
Understanding the origin and evolution of the Galaxy is one of the major goals of modern astronomy. The Galaxy stellar system is broadly composed of four distinct parts (although they likely overlap):  the bulge, thin disc, thick disc, and halo. The stellar kinematic and chemical abundance distributions are very important to characterize these components. Combining the BASS/DESI multi-epoch imaging observations with SDSS over the baseline of about 15 years, we will be able to measure proper motion of objects with $g < 22$ down to the level of 6 mas/year \citep{pen15}. This unique dataset will be combined with brighter objects from Gaia \citep{gai16}, allowing detailed studies of Galactic dynamics in the Galactic halo and a key probe to the evolution of our Galaxy.

\subsubsection{Searching satellite dwarf galaxies of the Galaxy}
The $\Lambda$-CDM model is highly successful in describing the observations of the large-scale structure and evolution of the Universe, but there remain tension between observations and simulations in describing structures on small scales. One of the main remaining problems is the missing satellite problem, namely the discrepancy between the number of predicted subhalos in N-body simulations and the observed number of Milky Way dwarfs. \citet{kly99} and \citet{moo99} first noticed that numerical simulations predicted much more subhaloes in galactic and cluster haloes. The number of satellites having circular velocities larger than Ursa-Minor and Draco were about 500 in simulations, while as it is well known that the MW has much less dSphs (9 bright dSphs, Sagittarius, the LMC and SMC).  The DESI imaging surveys will provide a new census of ultra-faint dwarf galaxies in the Milky Way Halo. We have calculated the volume that DESI imaging data will cover and the expected number of new discoveries based on \citet{tol08}. From Figure 9 and Table 4 in \citet{tol08}, our surveys are slightly shallower than the DES point, but we have an area of 14,000 deg$^2$ (2.8 times larger). So we are able to find dSphs within 250 kpc at $M_V < -4$, which will yield 8--30 new dwarfs. 

\subsubsection{Searching for substructures or tidal streams in the Galactic halo}
One of the key results from the SDSS is the discovery of ultra-faint dwarfs and tidal stream features in the Galactic halo, which revolutionized our understanding of the Galaxy assembly. However, at the SDSS depth, we can only effectively probe the inner halo. The BASS/DESI imaging data enable us to reach the Galactocentric distances of $~$40 kpc using main sequence stars \citep{jur08}, $~$80 kpc for turnoff stars selected using $g$ and $r$ colors \citep{bel08}, and $~$150 kpc for BHB stars if $u$-band data is available \citep{ruh11}. Theoretical models predict a dramatic increase in halo substructure at larger distance \citep{hel11}. Our deep imaging data will not only test this prediction, but also test the presence of additional satellite galaxies through their impact on the well known stream structures \citep{yoo11}.

\subsection{AGN evolution and high-redshift quasars}
The evolution of galaxies is tightly linked to the growth of their central super-massive black holes. Quasars may represent a key transitional phase in the evolution of massive galaxies \citep{hop06}. The SDSS has provided the most comprehensive optical census of AGN evolution so far \citep{ros13}, and opened a window for studying high-redshift quasar close to the end of cosmic reionization \citep{fan06,jia16}. The BASS/DESI imaging data will create an unprecedented sample of photometrically selected AGNs (a few million) over a wide range of redshift, AGN type, and luminosity. This AGN sample will allow us to study AGN clustering as a function of redshift which provides strong tests to the model of AGN evolution and black hole/galaxy co-evolution and to study how the overall SED of AGN correlate with redshift, luminosity, host galaxy properties and environment \citep{air12}. Because of the increased survey volume and depth, the BASS/DESI imaging surveys will be especially suitable for studying rarest types of AGNs and quasars. A key limiting factor of the SDSS quasar survey is the shallow $z$ band. The MzLS will reach about 3 magnitude deeper in $z$. If combined with the full-depth WISE data and the bluer deep BASS data, it will allow us to discover a large number of quasars at redshift up to 6.5, which maps the detailed luminosity function of faint quasars and provides key constraints on early growth of super-massive black holes.

\subsection{Photometric redshift and galaxy clusters}
Photometric redshifts have been widely used in studies of galaxy clusters and the large-scale structure of the Universe. The three BASS/DESI optical bands  and two WISE infrared bands (W1 and W2) provide multi-wavelength photometric data to estimate the redshifts of extragalactic sources in a very wide field. Photometric redshifts can be used for searching candidates of galaxy clusters whose photometric properties can be statistically studied. In addition, the SDSS has obtained the redshifts of 1.5 million massive galaxies. These galaxies are often the central and brightest galaxies in groups and clusters; current imaging data are not deep enough to detect  fainter members of the group/cluster. The BASS/DESI imaging will significantly improve stellar mass models for these galaxies and enable a sensitive search for cluster members around them. We will construct volume-limited cluster catalogs out to z$\sim$0.7, providing key cosmological probes \citep{bec07}.

\subsection{Spectroscopic Target Selection}
The DESI imaging surveys will be used for target selection of LRGs, ELGs, and QSOs for the DESI spectroscopic survey. DESI  spectroscopic sample an order of an magnitude larger than the SDSS. It will determine redshifts for about 18 million ELGs, 4 million LRGs, and 3 million QSOs. The survey is designed to probe the expansion history of the universe at $0 < z < 3.5$ using the baryon acoustic oscillation scale, and to map the dark matter and gravitational growth through redshift space distortions \citep{sch11,lev13}. DESI will also have revolutionary capabilities for broad investigations on the origin and evolution of galaxies, Galactic structure, rare-object discovery, and will produce one of the most complete high-z AGN catalogs to date \citep{pil12}. From the imaging surveys, we can also find a large number of rare classes objects, including halo and cool white dwarfs, cool L/T brown dwarfs, carbon stars, gravitationally-lensed quasars, high-z quasars etc. It will produce large and valuable datasets for spectroscopic follow-up observations. 

\section{Summary} \label{sec-summary}
BASS is a four-year collaborative program between NAOC and UA. It uses the 2.3 m Steward Observatory Bok telescope to image an area of about 5400 deg$^2$ of the Northern Galactic cap in the SDSS $g$ and DES $r$ bands. The expected 5$\sigma$ limiting AB magnitudes for point sources in $\sim$5 minutes are 24.0 and 23.4 mag, respectively. The BASS/DESI imaging data can be used for a wide range of astronomical researches from Galactic structures and external galaxies to high-redshift quasars and cosmology. Meanwhile, the data will be used for the DESI spectroscopic target selection. BASS started  observations in 2015 and will end in 2018.  About 41\% of the whole survey has been completed as of July 2016. Each BASS tile is covered by three passes or exposures that are dithered by 1/4 of the camera FoV. The three passes are taken under different weather conditions. This observing strategy has advantages for filling the CCD gaps, calculating accurate photometric solutions, and detecting variable objects. A number of ancillary tools are developed to improve  telescope pointing, focusing, and automating the observing. The exposure time calculator dynamically adjusts the exposure time according to real-time image quality in order to meet the depth requirement. We will have a new data release once a year, which will include newly observed data and update the data reduction pipeline. The early data produced by the SCUSS pipeline have been opened to the public and the first data release of BASS is ready for release.

\acknowledgments
The BASS is a collaborative program between the National Astronomical Observatories of Chinese Academy of Science and Steward Observatory of the University of Arizona. It is a key project of the Telescope Access Program (TAP), which has been funded by the National Astronomical Observatories of China, the Chinese Academy of Sciences (the Strategic Priority Research Program ``The Emergence of Cosmological Structures" Grant No. XDB09000000), and the Special Fund for Astronomy from the Ministry of Finance. The BASS is also supported by the External Cooperation Program of Chinese Academy of Sciences (Grant No. 114A11KYSB20160057) and Chinese National Natural Science Foundation (Grant No. 11433005). The BASS data release is based on the Chinese Virtual Observatory (China-VO).

\end{document}